# Nearly pure spin-valley sideband tunneling in silicene: effect of interplay of time periodic potential barrier and spin-valley-dependent Dirac mass

Ruanglak Jongchotinon[a], Bumned Soodchomshom[b]*

[a]Department of Mathematics, Faculty of Science, Kasetsart University, Bangkok 10900, Thailand

[b]Department of Physics, Faculty of Science, Kasetsart University, Bangkok 10900, Thailand

*Corresponding author
 Email address: Bumned@hotmail.com; fscibns@ku.ac.th

**Abstract**

We study massive Dirac fermion tunneling through time periodic potential in a silicene-based N/TP/N junction, where N's are normal silicene regions and TP is the time periodic potential barrier. The fermions would absorb or emit photons due to the presence of the Floquet sidebands created in the TP. The nearly perfect spin-valley-sideband filtering is predicted. Applying only the exchange field leads to just only the electron absorbing a single photon almost purely allowed to tunnel through the junction for the large electric field. Reversing direction of electric field can select spin of the allowed electron. In the case of applying only off-resonant circularly polarized light, just only the electron absorbing a single photon with spin up, is almost purely allowed to tunnel through the junction. The valley is also selected by reversing direction of the electric field. The controllable sideband channel may be applicable for sideband-based spin-valleytronics.

# 1. Introduction

Silicene, a monolayer of silicon atoms arranged in honeycomb lattice akin to graphene, has drawn much attention recently [1-3]. Graphitic form of silicon atoms has been first theoretically studied by Takeda and Shiraishi [4] and re-investigated by Guzman-Verri, et al [5]. After that, silicene has been confirmed to exist by growing on Ag(111) surface under ultrahigh vacuum conditions [6]. The first silicene-based field effect transistor (sFET) operated at room temperature has been fabricated in laboratory [7]. This is to show the potential of silicene as a 2-dimensional modern material for silicon-based advanced electronics [8]. High on/off ratio exceeding $10^5$ using sFET has been predicted by several works [9, 10]. Atomic structure of silicene is buckled while graphene is planar, due to the fact that the mixed orbital states $sp^2-sp^3$ when silicon atoms form as a graphitic structure [11]. Because of heavy atomic number, the intrinsic energy gap is generated by the presence of its intrinsic spin-orbit interaction (SOI) of about 3.9 meV [12]. The A and B sub-lattices are placed in different positions leading to gate tunable energy gap [13,14]. Not only energy gap is tunable by electric field but also exchange energy [15,16] and off-resonant circularly polarized light [17]. Unlike to that in graphene, fermions in the low energy regime in silicene are governed by the massive Dirac fermions with the gap-like mass of fermions in silicene being easily to be tuned by external forces [13-17]. The fermions in graphene are required specific substrates such as hBN [18], SiC[19,20] to open their energy gap. The mass term created by SOI and external forces would give rise to Berry curvature to get nonzero Chern numbers to turn silicene into a 2-dimensional topological material. Intriguingly, it exhibits rich topological phases classified by spin-valley dependent Chern numbers [21-23]. The magnetism and topological phase transition may be controlled by means of silicene grown on specific substrate to change its SOI [24]. Silicene would be a quantum spin Hall phase (QSH), a topological band with edge states, when perpendicular electric field is smaller than the critical value [21-23]. It undergoes quantum anomalous Hall phase (QAH), a topological band with edge states, when the magnetization is present [25]. The trivial band insulating phases, quantum valley (QVH) and quantum pseudo spin Hall (QPSH)phases, would occur when applying electric field and off-resonant circularly polarized light [15, 26-28].

The emergence of spin-valley dependent massive Dirac fermions in silicene would give rise to the bridge between high energy and condensed matter physics, similar to graphene [29]. Quantum behavior in silicene may be considered as a system like 2-dimensional quantum electrodynamics. According to the new behavior of electron unlike Schrödinger type, the photon-assisted tunneling in graphene has been drawn much attention [30-42]. This photon-assisted relativistic electron tunneling has been predicted to lead to novel behavior different from that in Schrödinger electron system [43]. Tunneling processes through scalar time periodic potential [30-32,34-36,38,39,41,42] and polarized electromagnetic wave [33,37,40] applied in to the barriers have been studied where the effect of Dirac fermions nature has been focused. The time-dependent sinusoidal potential with frequency f would create photon with energy $E_{ph}=\hbar 2\pi f=\hbar\omega$, due to the presence of the Floquet quasi energy Eigen states [36,43]. Electron with energy E is injected into the barrier and then leaves the barrier with the energy of $E+(-)n\hbar\omega$ (n=0,1,2,...) due to absorbing (emitting) n-photons [38, 43]. The tunneling through time periodic potential in graphene shows the transmission oscillating due to Klein tunneling and coupling with photon in case of massless fermions [30,31,34-38,40,41]. The Klein tunneling was suppressed by the irradiation of a strong laser field [33,37,40]. Massive fermions have been also found to suppress Klein tunneling coupling with photon [32,42]. Due to the intriguing properties of silicene, the carriers are governed by the fermions with Dirac mass controlled by external forces such as perpendicular electric field and staggered exchange energy [13-17]. This is quite different nature from those in graphene. Since silicene is a topological phase that may change topological phases by varying external force [21-23], the tunneling of electrons in silicene through the time periodic potential is interesting.

In this letter, we study spin-valley dependent tunneling behavior of controllable massive Dirac fermion in silicene-based N/TP/N junction, where N's are normal silicene regions and TP is the time periodic potential barrier. The perpendicular electric field, the staggered exchange energy, off-resonant circularly polarized light and gate potential are applied into the barrier. The time periodic potential leads to the Floquet quasi energy Eigen states [36,43] to create photons coupling with tunneling electron. Mass of Dirac fermions in the barrier may be controlled by varying the electric field, exchange field and off-resonant circularly polarized light. We study the transmission probabilities of central band and side bands. In this work, we focus on spin-valley sideband filtering effect that may be applicable for sideband-based spin-valleytronics devices.

## 2. Model and Scattering process

In our model, the silicene-based N/TP/N junction is depicted in Fig.1a. The barrier region TP is assumed to be under the influence of the time periodic potential, electric field $E_z$., exchange field and off-resonant circularly polarized light. In the pristine silicene, the system is governed by four species of Dirac fermion in k and k' valley with spin $\uparrow(\downarrow)$. In this work, the tunneling property is to focus on the low energy limit. The effect of small strength of Rashba spin orbit interaction may be neglected [12, 15-17]. In the low energy limit, the carriers in silicene may be described by the massive relativistic behavior. In the absence of time periodic potential and external forces for $|x| > L$, the wave function may obey the massive Dirac equation given by [12, 15-17,21-28]

$$(v_F \eta \hat{p}_x \sigma_x + v_F \hat{p}_y \sigma_y + \eta s \Delta_{so} \sigma_z) \psi_{\eta s}(x,y,t) = i\hbar \partial_t \psi_{\eta s}(x,y,t),$$

(1)

where, $\hat{p}_{x(y)} = -i\hbar \partial_{x(y)}$ and the Pauli spin matrices acting on sublattice pseudo spin state are $\sigma_{x,y,z}$. The notations $\eta = 1(-1)$ denote the electron in the k- ( k'-) valley with $s = 1(-1)$ representing spin $\uparrow(\downarrow)$. In silicene, the intrinsic spin orbit interaction and the Fermi velocity are $\Delta_{so} \cong 3.9$ meV and $v_F \cong 5.5 \times 10^5$ m/s, respectively [12].

In the TP-region for $|x| \leq L$, the influence of time periodic potential $V(t) = V_0 + V_1 \cos \omega t$ [30-32,34-36,38,39,41,42] and the external fields are taken into account for photon creation and spin-valley-dependent gap opening, respectively. The motion of electron may be described using the wave equation of the form

$$(v_F \eta \hat{p}_x \sigma_x + v_F \hat{p}_y \sigma_y + \Delta_{\eta s} \sigma_z + V(t)) \psi'_{\eta s}(x,y,t) = i\hbar \partial_t \psi'_{\eta s}(x,y,t),$$

(2)

where $m = \Delta_{\eta s}/v_F^2 = (\eta s \Delta_{so} + \Delta_z + \eta \Delta_\gamma + s \Delta_M)/v_F^2$ is the spin-valley dependent gap-like Dirac mass. The electric field induced gap $\Delta_z = eE_z D$ with $D = 0.23 \overset{o}{A}$ being buckling parameter is generated due to its buckled atomic structure [13-15]. The valley-dependent gap $\eta \Delta_\gamma$ induced by the off-resonant circularly polarized light with vector potential of $\vec{A} = \Lambda < \pm \sin\gamma t, \cos\gamma t, 0 >$, where $\Lambda$ and $\gamma$ are the amplitude and the frequency, respectively. The strength of the interaction with right (left) circulation is described by the parameter $\Delta_\gamma = +(-) v_F^2 e^2 a^2 \Lambda^2 / (\hbar\Omega)$, where $a = 3.86 \mathring{A}$ is the lattice constant [17,44,45]. The staggered magnetic insulators may create the spin-dependent gap term $s\Delta_M$ because of the A and B sublattices having opposite sign of the exchange energy [15,16]. The spin-valley dependent relativistic mass of the Dirac fermions in silicene would be given by the relation of the form $m_{\eta s} = \Delta_{\eta s} / v_F^2$. The solution to eq.(2) may be assumed as

$$\psi'_{\eta s}(x, y, t) = e^{-iEt/\hbar + ik_y y} \phi(x, t), \qquad (3)$$

where E is the Floquet quasi energy and $\phi(x,t) = \phi(x, t+T)$ with T is the period of time periodic potential. The wave vector $\theta$ in the y direction $k_y = \sin\theta \sqrt{E^2 - \Delta_{so}^2}/\hbar v_F$ is the conservation component where $\theta$ is the incident angle. We use the method of separation of variables by substituting $\phi(x,t) = F(t)(\varphi_A(x), \varphi_B(x))^T$ into eqs.(3) and (2). The eq.(2) is thus split into two parts, time and space components, as given by

$$i\hbar \partial_t f(t) - (V_1 \cos\omega t) f(t) = E f(t), \qquad (4a)$$

$$\begin{pmatrix} \Delta_{\eta s} + V_0 & v_F \eta \hat{p}_x - i\hbar v_F k_y \\ v_F \eta \hat{p}_x + i\hbar v_F k_y & -\Delta_{\eta s} + V_0 \end{pmatrix} \begin{pmatrix} \varphi_A(x) \\ \varphi_B(x) \end{pmatrix} = E \begin{pmatrix} \varphi_A(x) \\ \varphi_B(x) \end{pmatrix}. \qquad (4b)$$

By the help of the generating function of the Bessel function, the time dependent solution to eq.4a may be given as $f(t) \sim e^{-iEt/\hbar} e^{-i\alpha \sin\omega t} = e^{-iEt/\hbar} \sum_{n=-\infty}^{=+\infty} J_n(\alpha) e^{-in\omega t}$ where $J_n(\alpha)$ is the nth-order Bessel function. Therefore, the solution to eqs.(4a) and (4b) describing the wave function inside the barrier may be obtained as

$$\psi_2(x \leq |L|) = \sum_{n=-\infty}^{\infty} \sum_{l=-\infty}^{+\infty} [a_l \begin{pmatrix} 1 \\ \frac{(E - V_0 + l\hbar\omega) - \Delta_{\eta s}}{\eta \hbar v_F q_{x,l} - i\hbar v_F k_y} \end{pmatrix} e^{iq_{x,l} x} ] J_n(\frac{V_1}{\hbar\omega}) e^{ik_y y - i\frac{(E + n\hbar\omega + l\hbar\omega)}{\hbar} t}$$

$$+ \sum_{n=-\infty}^{\infty} \sum_{l=-\infty}^{+\infty} [b_l \begin{pmatrix} 1 \\ \frac{(E - V_0 + l\hbar\omega) - \Delta_{\eta s}}{-\eta \hbar v_F q_{x,l} - i\hbar v_F k_y} \end{pmatrix} e^{-iq_{x,l} x} ] J_n(\frac{V_1}{\hbar\omega}) e^{ik_y y - i\frac{(E + n\hbar\omega + l\hbar\omega)}{\hbar} t},$$

(5)

where $q_{x,l} = \sqrt{(E - V_0 + l\hbar\omega)^2 - \Delta_{\eta s}^2 - (\hbar v_F k_y)^2}/\hbar v_F$ and $a(b)_l$ is amplitude of wave function.

In the left-N region, the injected electron wave function $\psi^i$ must obey eq.(1), given by

$$\psi^i = [\begin{pmatrix} 1 \\ \frac{E - \eta s \Delta_{so}}{\eta \hbar v_F k_{x,0} - i\hbar v_F k_y} \end{pmatrix} e^{ik_{x,o}x}] e^{ik_y y - i\frac{E}{\hbar}t}, \qquad (6)$$

where $k_{x,0} = \sqrt{(E)^2 - \Delta_{so}^2 - (\hbar v_F k_{//})^2}/\hbar v_F$. Due to the oscillating potential barrier creating a photon with energy $\hbar\omega$, the reflected electron can thus exchange energy in units of $\hbar\omega$. The spectrum energy of reflected wave $\psi_1^r$ and transmitted wave $\psi_1^t$ coupling with photons may be given by $E \to E + l\hbar\omega$. Positive (negative) sign of l describe absorbing(emitting) $|l|$ numbers of photons. The wave function in the left-N region may be as $\psi_1(x < L) = \psi^i + \psi^r$, to get

$$\psi_1 = [\begin{pmatrix} 1 \\ \frac{E - \eta s \Delta_{so}}{\eta \hbar v_F k_{x,0} - i\hbar v_F k_y} \end{pmatrix} e^{ik_{x,o}x} + \sum_{l=-\infty}^{+\infty} r_l \begin{pmatrix} 1 \\ \frac{(E + l\hbar\omega) - \eta s \Delta_{so}}{-\eta \lambda_l \hbar v_F k_{x,l} - i\hbar v_F k_y} \end{pmatrix} e^{-\lambda_l i k_{x,l} x - il\omega t}] e^{ik_y y - i\frac{E}{\hbar}t},$$

(7)

where $k_{x,l} = \sqrt{(E + l\hbar\omega)^2 - \Delta_{so}^2 - (\hbar v_F k_y)^2}/\hbar v_F$. $\lambda_l = \text{sgn}(E + l\hbar\omega)$ for $k_{x,l}$ being real and $\lambda_l = -1$ for $k_{x,l}$ being imaginary. When In the right-N region, the wave function is described by the transmitted part $\psi_3(x > L) = \psi^t$, as obtained as

$$\psi_3 = \sum_{l=-\infty}^{+\infty} t_l \begin{pmatrix} 1 \\ \frac{(E + l\hbar\omega) - \eta s \Delta_{so}}{\eta \lambda_l \hbar v_F k_{x,l} - i\hbar v_F k_y} \end{pmatrix} e^{\lambda_l i k_{x,l} x + i k_y y - i\frac{E + l\hbar\omega}{\hbar}t}.$$

(8)

The coefficients $r_l$ and $t_l$ belong to the reflected and transmitted wave that interchange energy of $l\hbar\omega$.

The coefficients in eqs.(5), (7) and (8) may be determined by using the orthogonality of $(1/2\pi)\int_{-\pi}^{\pi} e^{imt} e^{-im't} dt = \delta_{mm'}$ and the boundary conditions at the interfaces $\psi_1(0) = \psi_2(0)$ and $\psi_2(L) = \psi_3(L)$. By doing this, it is given the linear system of equations for $n, l = 0, \pm 1, \pm 2, ...$, obtained as

$$\delta_{n0} \begin{pmatrix} 1 \\ \frac{E - \eta s \Delta_{so}}{\eta \hbar v_F k_{x,0} - i\hbar v_F k_y} \end{pmatrix} + r_n \begin{pmatrix} 1 \\ \frac{(E + n\hbar\omega) - \eta s \Delta_{so}}{-\eta \lambda_n \hbar v_F k_{x,n} - i\hbar v_F k_y} \end{pmatrix} =$$
$$\sum_{l=-\infty}^{\infty} \left[ a_l \begin{pmatrix} 1 \\ \frac{(E - V_0 + l\hbar\omega) - \eta s \Delta_{\eta s}}{\eta \hbar v_F q_{x,l} - i\hbar v_F k_y} \end{pmatrix} + b_l \begin{pmatrix} 1 \\ \frac{(E - V_0 + l\hbar\omega) - \eta s \Delta_{\eta s}}{-\eta \hbar v_F q_{x,l} - i\hbar v_F k_y} \end{pmatrix} J_{n-l}(\frac{V_1}{\hbar\omega}) \right]$$

(9)

$$t_n \begin{pmatrix} 1 \\ \dfrac{(E+n\hbar\omega)-\eta s\Delta_{so}}{\eta\lambda_n\hbar v_F k_{x,n}-i\hbar v_F k_y} \end{pmatrix} e^{\lambda_n i k_{x,n} L} =$$

$$\sum_{l=-\infty}^{\infty}\left[ a_l \begin{pmatrix} 1 \\ \dfrac{(E-V_0+l\hbar\omega)-\Delta_{\eta s}}{\eta\hbar v_F q_{x,l}-i\hbar v_F k_y} \end{pmatrix} e^{iq_{x,l}L} + b_l \begin{pmatrix} 1 \\ \dfrac{(E-V_0+l\hbar\omega)-\Delta_{\eta s}}{-\eta\hbar v_F q_{x,l}-i\hbar v_F k_y} \end{pmatrix} e^{-iq_{x,l}L} \right] J_{n-l}(\dfrac{V_1}{\hbar\omega})$$

(10)

### 3. Spin-valley-sideband-dependent transmission formulae

The reflection and transmission of the junction may be calculated via the probability current density of the Dirac type $J_x = \langle \psi | v_F \sigma_x | \psi \rangle$ where $J_i = \langle \psi^i | v_F \sigma_x | \psi^i \rangle$, $J_t = \langle \psi^t | v_F \sigma_x | \psi^t \rangle$ and $J_R = \langle \psi^r | v_F \sigma_x | \psi^r \rangle$. From eqs.(7) and (8), we may have $\psi^r = \sum_{l=-\infty}^{\infty} \varphi^r_{l\eta s}$ and $\psi^t = \sum_{l=-\infty}^{\infty} \varphi^t_{l\eta s}$. The spin-valley-sideband dependent transmission and reflection are therefore respectively given as

$$T_{l\eta s} = \dfrac{J_{t,l\eta s}}{J_i} = \dfrac{\langle \varphi^t_{l\eta s} | v_F \sigma_x | \varphi^t_{l\eta s} \rangle}{J_i} = t_l^* t_l \dfrac{\left( \dfrac{(E+l\hbar\omega)-\eta s\Delta_{so}}{\eta\lambda_l \hbar v_F k_{x,l} - i\hbar v_F k_y} + \dfrac{(E+l\hbar\omega)-\eta s\Delta_{so}}{\eta\lambda_l \hbar v_F k_{x,l} + i\hbar v_F k_y} \right)}{\left( \dfrac{E-\eta s\Delta_{so}}{\eta\hbar v_F k_{x,0} - i\hbar v_F k_y} + \dfrac{E-\eta s\Delta_{so}}{\eta\hbar v_F k_{x,0} + i\hbar v_F k_y} \right)},$$

(11)

and

$$R_{l\eta s} = \dfrac{J_{r,l\eta s}}{J_i} = \dfrac{\langle \varphi^r_{l\eta s} | v_F \sigma_x | \varphi^r_{l\eta s} \rangle}{J_i} = r_l^* r_l \dfrac{\left( \dfrac{(E+l\hbar\omega)-\eta s\Delta_{so}}{-\eta\lambda_l \hbar v_F k_{x,l} - i\hbar v_F k_y} + \dfrac{(E+l\hbar\omega)-\eta s\Delta_{so}}{-\eta\lambda_l \hbar v_F k_{x,l} + i\hbar v_F k_y} \right)}{\left( \dfrac{E-\eta s\Delta_{so}}{\eta\hbar v_F k_{x,0} - i\hbar v_F k_y} + \dfrac{E-\eta s\Delta_{so}}{\eta\hbar v_F k_{x,0} + i\hbar v_F k_y} \right)}.$$

(12)

Based on the continuity condition, we may get $1 = \sum_{l=-\infty}^{\infty} R_{l\eta s} + \sum_{l=-\infty}^{\infty} T_{l\eta s}$ because $J_i = J_R + J_t$. The spin-valley dependent transmission may be defined as

$$T_{\eta s} = \sum_{l=-\infty}^{\infty} T_{l\eta s}. \qquad (13)$$

The total transmission may be given as the summation of all spin-valley dependent transmission, given by

$$T_{total} = T_{k\uparrow} + T_{k\downarrow} + T_{k'\uparrow} + T_{k'\downarrow} \qquad (14)$$

### 4. Numerical result and discussion

In the numerical result, we study the transmission of tunneling electron in the regime that $V_1/\hbar\omega \ll 1$. Beside the incident channel of energy E, three Floquet sidebands above and below E are considered, $E_n = E, E\pm\hbar\omega, E\pm 2\hbar\omega, ..., E\pm N\hbar\omega$, where N=3. In the numerical result, we should have $N > V_1/\hbar\omega$ [46, 47]. The central band $T_{0\eta s}$, the first $T_{\pm 1\eta s}$, second $T_{\pm 2\eta s}$ and the third $T_{\pm 3\eta s}$ sidebands will be numerically shown. The investigation would show the spin-valley dependent transmissions of the electron that absorbs(emits) zero, one two and three photons. The transmissions as a function of angle of incidence are firstly investigated in Fig.2. The frequency $f = 2000$ GHz and $V_1 = 5$ meV would give rise to $\hbar\omega \cong 8.27$ meV and $N(=3) > V_1/\hbar\omega \cong 0.6$. In this study, only the electric field and staggered exchange energy are applied. The effect of light is canceled. It is found that the spin-valley transmission in each sideband depends on spin, valley and sideband. The spin-valley dependent effect is due to the fact that the gap-like mass of Fermions with different spins and valleys are different ie.,

$$E_{gap,\eta s} = 2|\Delta_{\eta s}| = 2|\eta s \Delta_{so} + \Delta_z + \eta \Delta_\gamma + s \Delta_M|, \quad (15)$$

to get different energy gap for fermions $|\Delta_{k\uparrow}| = 11.4$ meV, $|\Delta_{k\downarrow}| = 6.4$ meV, $|\Delta_{k'\uparrow}| = 3.6$ meV and $|\Delta_{k'\downarrow}| = 1.4$ meV. The sideband-dependent effect is generated because of the difference in the energy level of electron in the TP-barrier $E - V_0 \pm l\hbar$ and in N regions $E \pm l\hbar$ for reflected and transmitted electronic waves, as depicted Fig.1(c). In Figs2b and 2e, $T_{1\eta s}$ is found inside the larger region than $T_{-1\eta s}$. This is because the sideband transmission may be suppressed beyond the critical angle $T_{l\eta s}(\theta \geq |\theta_{c,l}|) = 0$, calculated via the conservation of the parallel component, as given by

$$\sin\theta_{c,l} = \frac{\sqrt{(E+l\hbar\omega)^2 - \Delta_{so}^2}}{\sqrt{E^2 - \Delta_{so}^2}}. \quad (16)$$

Also, because of small $V_1/\hbar\omega \cong 0.6$, the transmission for the third sidebands are then almost perfectly suppressed (see Figs.2d and 2g). In this junction, the spin-valley transmission may be approximately given by $T_{\eta s} \cong \sum_{l=-2}^{+2} T_{l\eta s}$ (see Fig.2h). If $V_1/\hbar\omega$ is large, the transmission may be dominated by including the large number of l. The presence of spin-valley-sideband dependent transmission is due to the interplay of the presence spin-valley-dependent mass and time periodic potential in silicene. This is quite different from that in graphene system [30,31,34-38,40,41], because sublattice symmetry breaking mass of fermions in graphene is not spin-valley dependent [18-20,32,42].

We next study the transmission as a function of electric field under the effect of applying staggered exchange energy $\Delta_M = 2$ meV. The effect of light is still canceled. In Fig.3, we have $E = V_0$. The Fermi level in the barrier is zero or lies at the Dirac point. In this case, split of the energy level $E_l = E - V_0 \pm l\hbar = \pm l\hbar$ about the Dirac point is symmetric to induce the sideband-independent effect in the case of the same Floquet sideband $T_{+1\eta s} = T_{-1\eta s}, T_{+2\eta s} = T_{-2\eta s}$ and $T_{+3\eta s} = T_{-3\eta s}$, as seen in Figs. 3(b) to 3(g). The spin-valley transmission is split into four peaks at the electric field calculated by the condition of $\Delta_{z,\eta s} = -(\eta s \Delta_{so} + \eta \Delta_\gamma + s \Delta_M)$ for $T_{l\eta s}$ [16,22]. The peaks of $T_{lk\uparrow}, T_{lk\downarrow}, T_{lk'\uparrow}$ and $T_{lk'\downarrow}$ are thus found at $-5.9$ meV, $+5.9$ meV, $+1.9$ meV and $-1.9$ meV, respectively, as seen in Fig3. The spin-valley filtering like this in Fig.3(a) has been reported in a staggered exchange field induced gap in silicene junctions without time periodic potential barrier [16,22]. In this work, the spin-valley filter behavior appears in all sidebands due to the presence of Floquet quasi Eigen energies.

The nearly pure spin-valley-sideband tunneling may be predicted when $E \neq V_0$, see Fig.4. The Fermi level of central band is not zero $E - V_0 \neq 0$. In this case we take $\Delta_M = 2\text{meV}$ and $\Delta_\gamma = 0$ (no applied light) to get

$$\Delta_{\eta s}(\Delta_M \neq, \Delta_\gamma = 0) = \eta s \Delta_{so} + \Delta_z + s \Delta_M. \qquad (17)$$

It is found that the shape of all spin-valley-sideband transmission are different. It is very interesting that almost all spin-valley-sideband transmissions are suppressed completely for $|\Delta_Z| = 20\text{meV}$ except for $T_{1k\uparrow} \neq 0$ and $T_{1k\downarrow} \neq 0$ (see Fig.4(b)). The junction is almost governed by electron in the 1$^{st}$-sideband of the k-valley with spin $\uparrow$ tunnels through the junction when $\Delta_Z < -20\text{meV}$. The junction is also almost governed by electron in the 1$^{st}$-sideband of the k-valley with spin $\downarrow$ tunnels through the junction when $\Delta_Z > +20\text{meV}$. This junction shows the excellent potential to **switch the spin of electron** in the 1$^{st}$-sideband of the k-valley by reversing direction of electric field $\Delta_Z \pm 20\text{meV}$.

It is similar to the result given in Fig.4. When the exchange field is replaced by the off-resonant circularly polarized light $\Delta_\gamma = 2\text{meV}$ and $\Delta_M = 0$, the $T_{1\eta s}$ in Fig.5 are similar to the behavior of the $T_{1s\eta}$ in Fig.4. This is due to the fact that the gap

$$\Delta_{\eta s}(\Delta_M = 0, \Delta_\gamma \neq 0) = \eta s \Delta_{so} + \Delta_z + \eta \Delta_\gamma, \qquad (18)$$

in the eq.(18) is the form of the interchange $s \leftrightarrow \eta$ in eq.(17). The junction is almost governed by electron in the 1$^{st}$-sideband of the k-valley with spin $\uparrow$ tunnels through the junction when $\Delta_Z < -20\text{meV}$. The junction is also almost governed by electron in the 1$^{st}$-sideband of the $k'$-valley with spin $\uparrow$ tunnels through the junction when $\Delta_Z > +20\text{meV}$. This junction shows the excellent potential to **switch the valley of electron** in the 1$^{st}$-sideband with spin $\uparrow$ by reversing direction of electric field $\Delta_Z \pm 20\text{meV}$. Finally, the ratio of spin-valley-sideband transmission per the total transmission is investigated in Fig.6. It is found that the $T_{1k\uparrow}/T_{total}$ and $T_{1k\downarrow}/T_{total}$ increase as the frequency increases, see Figs.6(b) and 6(d). The numbers of the peaks of $T_{1k\uparrow(\downarrow)}$ for $|\Delta_Z| < 20\text{meV}$ also increases when increasing f, see Figs.6(b) and 6(d). At $f = 10000\,\text{GHz}$, the result shows $T_{1k\uparrow(\downarrow)}/T_{total} \cong 0.98$. This prediction confirms that the tunneling may be almost carried by one channel of spin-valley-sideband electron. Transmission may be almost completely governed by the electron which absorbs one photon in the barrier.

## 5. Summary and conclusion

We have investigated tunneling of Dirac fermions through silicene junction with time periodic potential barrier. In the barrier, perpendicular electric field, exchange field and off-resonant circularly polarized light have been assumed to be applied into the barrier. Due to the presence of Floquet quasi Eigen energies and controllable spin-valley-dependent Dirac mass of fermions in the barrier, the junction has shown great potential for application of sideband-based spin-valleytronics. When applying only the exchange field into the barrier, the total transmission of the junction is almost governed by the electron absorbing one photon (or electron in the 1$^{st}$-Froquet sideband) in the k valley. The spin ↑(↓) may be controlled by reversing the direction of the electric field −(+). Interestingly, the total transmission of the junction is almost governed by the electron absorbing one photon with spin up in the valley k(k′) when the only off-resonant circularly polarized light is applied. The selected valley k(k′) may be controlled by the direction of the direction of electric field −(+). Nearly pure spin and pure valley of the electron absorbing one photon tunneling in the silicene-based junction predicted in the present work has been revealed that the spin-valley dependent Dirac mass of fermions in silicene is the key behind this intriguing behavior.

**Acknowledgment**
This work is supported by the Kasetsart University Research and Development Institute (KURDI).

**References**

[1] A. Molle, et al., Chem. Soc. Rev. **47**(2018)6370
[2] H. Oughaddou, et al., Progress in Surface Science **90**(2015)46
[3] J. Zhao et al., Progress in Materials Science **83** (2016) 24
[4] K. Takeda, K. Shiraishi, Phys. Rev. **B** 50(1994) 14916
[5] G. G. Guzman-Verri, L. C. Lew Yan Voon, Phys. Rev.B**76** (2007) 075131
[6] P. Vogt, et al., Phys. Rev. Lett. **108** (2012) 155501
[7] L. Tao, et al., Nat. Nanotech.**10** (2015) 227
[8] Q. Ru-Ge, et al.,Chin. Phys. B **24** (2015) 088105
[9] M. P. Lima, et al., IEEE Electron Device Letters **39** (2018) 1258
[10] M. Vali, et al., Journal of Computational Electronics **15** (2016)138
[11] E. Cinquanta, *et al.* J. Phys. Chem. C **117** (2013)16719
[12] C.-C. Liu, H. Jiang, Y. Yao, Phys. Rev. B **84** (2011)195430
[13] N.D. Drummond, V. Zolyomi, V.I. Fal'ko, Phys. Rev. B **85** (2012)075423
[14] M. Houssa, *et al.*, Phys. Chem. Chem. Phys. **15** (2013) 3702
[15] M. Ezawa, Phys. Rev. B **87** (2013) 155415
[16] B. Soodchomshom, J. Appl. Phys.**115** (2014) 023706
[17] M. Ezawa, Phys. Rev. Lett. **110** (2013) 026603
[18] G. Giovannetti, et.al., Phys.Rev. B **76** (2007)073703
[19] M. S. Nevius et.al., Phys.Rev.Lett. **115** (2015)136802
[20] S.Y. Zhou, *et al.*Nat. Mater. **6** (2007) 770
[21] M. Ezawa, J. Phys. Soc. Jpn. **84** (2015) 121003
[22] W. Prarokijjak, B. Soodchomshom, J. Magn. Magn Mater. **452** (2018) 407
[23] M. Ezawa, Phys. Rev. B **88** (2013) 161406(R)
[24] K. Yang, et al., Nanoscale **10** (2018) 14667
[25] M. Ezawa**,** Phys. Rev. Lett. 109 (2012) 055502
[26] H. Bao, *et al.*, J. Appl. Phys., **121** (2017) 205106
[27] Y. Mohammadi, B.A. Nia, Superlattice. Microst. **96** (2016) 259
[28] W. Prarokijjak, B. Soodchomshom, Physica E**114** (2019) 113584
[29] M.I. Katsnelson, K.S. Novoselov, Solid State Communications **143** (2007) 3–13

[30] B. Trauzettel, et al., Phys. Rev. B **75** (2007) 035305
[31] M. Ahsan Zeb, et al., Phys. Rev. B **78** (2008) 165420
[32] Z.-Z. Cao, et al., Physics Letters A **375** (2011) 4065
[33] C. Sinha, R. Biswas, Appl. Phys. Lett. **100** (2012)183107

[34] W.-T. Lu et al., J. Appl. Phys. **111** (2012)103717
[35] S. E. Savel'ev, et al., Phys. Lett. **109**(2012) 226602
[36] H. P. Ojeda-Collado, C. Rodríguez-Castellanos, Appl. Phys. Lett. **103**(2013) 033110
[37] R. Biswas, and C. Sinha, J. Appl. Phys. **114**(2013) 183706
[38] A. Jellal, et al., Eur. Phys. J. B **87** (2014)123
[39] H. Chen, Physica B **456** (2015)167
[40] R. Biswas, et al., Physica E **84** (2016) 235
[41] R.Biswas, et al., Physics Letters A **381** (2017) 1582
[42] W. Yan, Physica B **504** (2017) 23
[43] P. E. Tien, J. P. Gordon, Phys. Rev.**129**(1963)647
[44] M. Tahir, U. Schwingenschlögl, Eur. Phys. J. B **88** (2015) 285.
[45] A. López, A. Scholz, B. Santos, J. Schliemann, Phys. Rev. B **91** (2015) 125105.
[46] G. P. Berman, E. N. Bulgakov, D. K. Campbell, and A. F. Sadreev, Physica B **225** (1996)
[47] W. Li, L. E. Reichl, Phys.Rev. B **60** (1999)15732

### Figures captions

**Figure 1** schematic illustration of the studied model (a) scattering process(b) and the energy level of each sideband in the N and TP regions (c). In the barrier, we assume that the electric field, staggered exchange fields from magnetic insulators and off-resonant circularly polarized light are applied, in order to control the mass of fermions. The time periodic potential $V_0 + V_1 \cos\omega t$ is applied into the barrier via the gates.

**Figure 2** the transmission $T_{l\eta s}$ as a function of angle of incidence in the case of the presence of electric, exchange fields $\Delta_z = 2.5\text{meV}$, $\Delta_M = 5\text{meV}$ and $\Delta_\gamma = 0$. The transmission is found to depend on sideband, spin and valley.

**Figure 3** the transmission $T_{l\eta s}$ as a function of electric field gap $\Delta_z$ for applying only staggered exchange energy when $E = V_0$. In this case, the nearly pure sideband-spin-valley electron tunneling process does not take place.

**Figure 4** the transmission $T_{l\eta s}$ as a function of the electric field gap $\Delta_z$ for applying only the staggered exchange energy $\Delta_M \neq 0$ and $\Delta_\gamma = 0$, when $E \neq V_0$. The electron with $k,\uparrow$ and $k,\downarrow$ in the 1$^{st}$ Floquet sideband are the main carrier when $\Delta_z < -20\text{meV}$ and $\Delta_z > -20\text{meV}$, respectively.

**Figure 5** the transmission $T_{l\eta s}$ as a function of electric field gap $\Delta_z$ for applying only off-resonant circularly polarized light $\Delta_M = 0$ and $\Delta_\gamma \neq 0$, when $E \neq V_0$. The electron with $k,\uparrow$ and $k',\uparrow$ in the 1$^{st}$ Floquet sideband are the main carrier when $\Delta_z < -20\text{meV}$ and $\Delta_z > -20\text{meV}$, respectively.

**Figure 6** the transmission of sideband l=1 is investigated in (a) and (b). The ratio of transmission per total transmission of the electron with $k,\uparrow$ and $k,\downarrow$ in the 1$^{st}$ Floquet sideband is investigated in (b) and (d). The transmission ratio of the electron with $k,\uparrow$ and $k,\downarrow$ may approach one when increasing the frequency of time periodic potential to get to get the nearly pure spin-valley-sideband electron tunneling.

(1a)

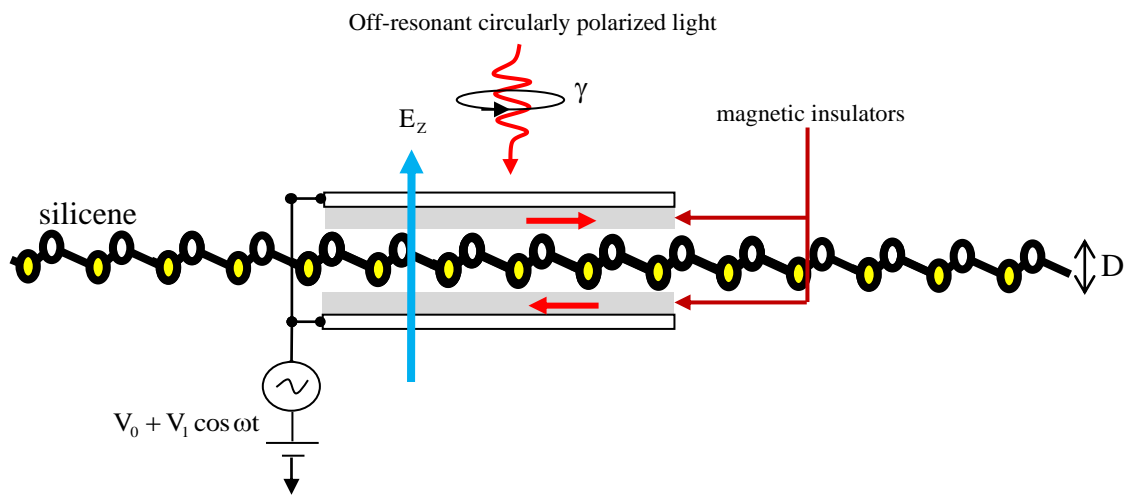

(1b)

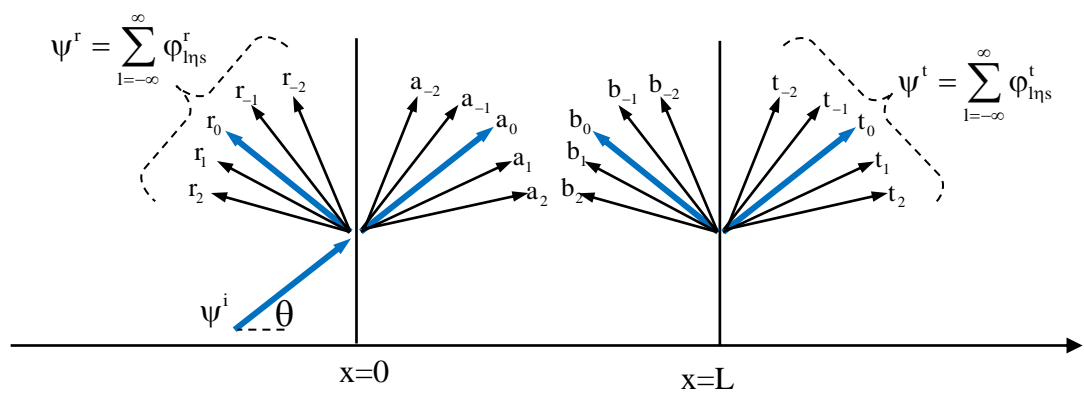

(1c)

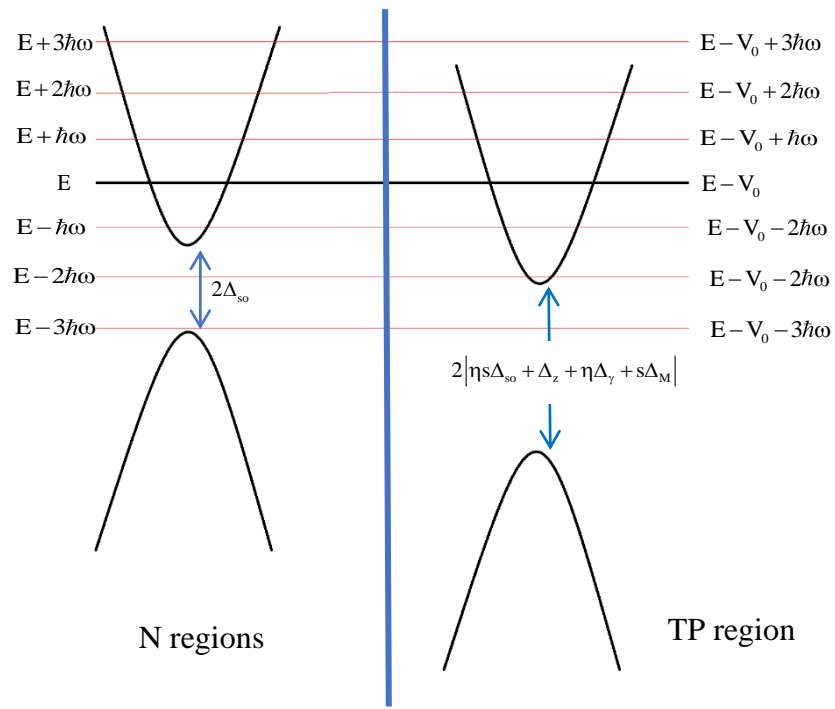

**Figure 1**

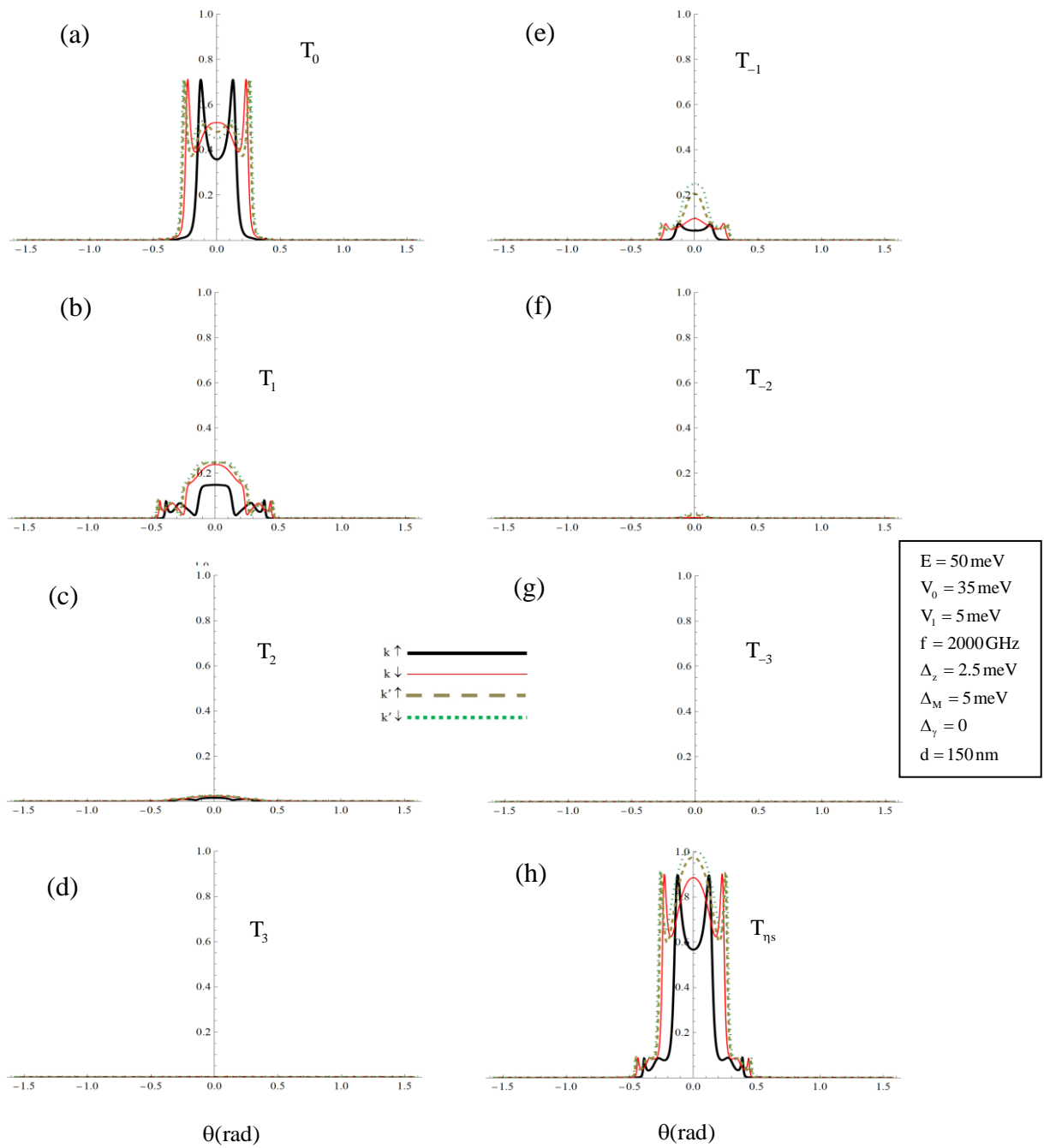

Figure 2

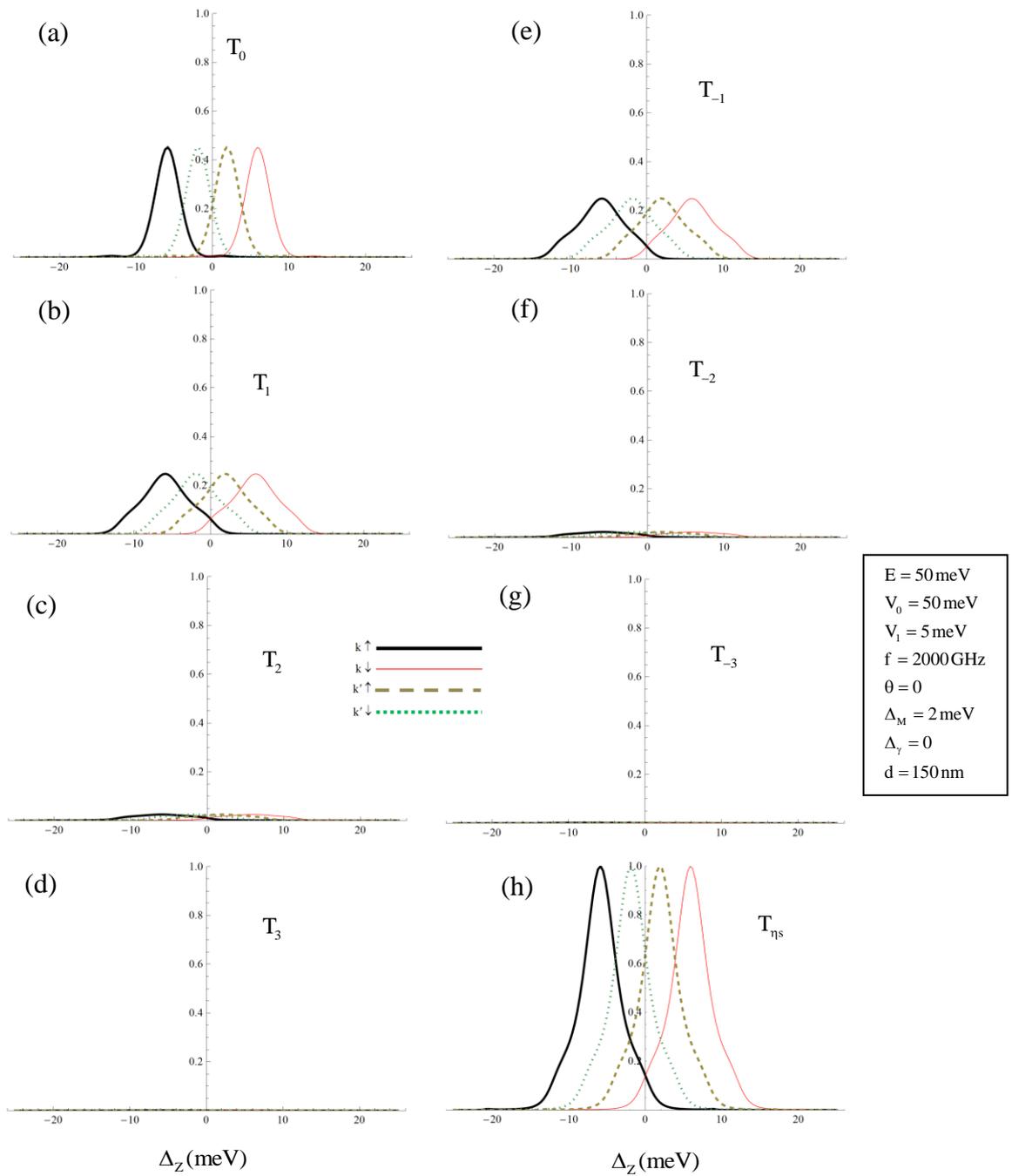

**Figure 3**

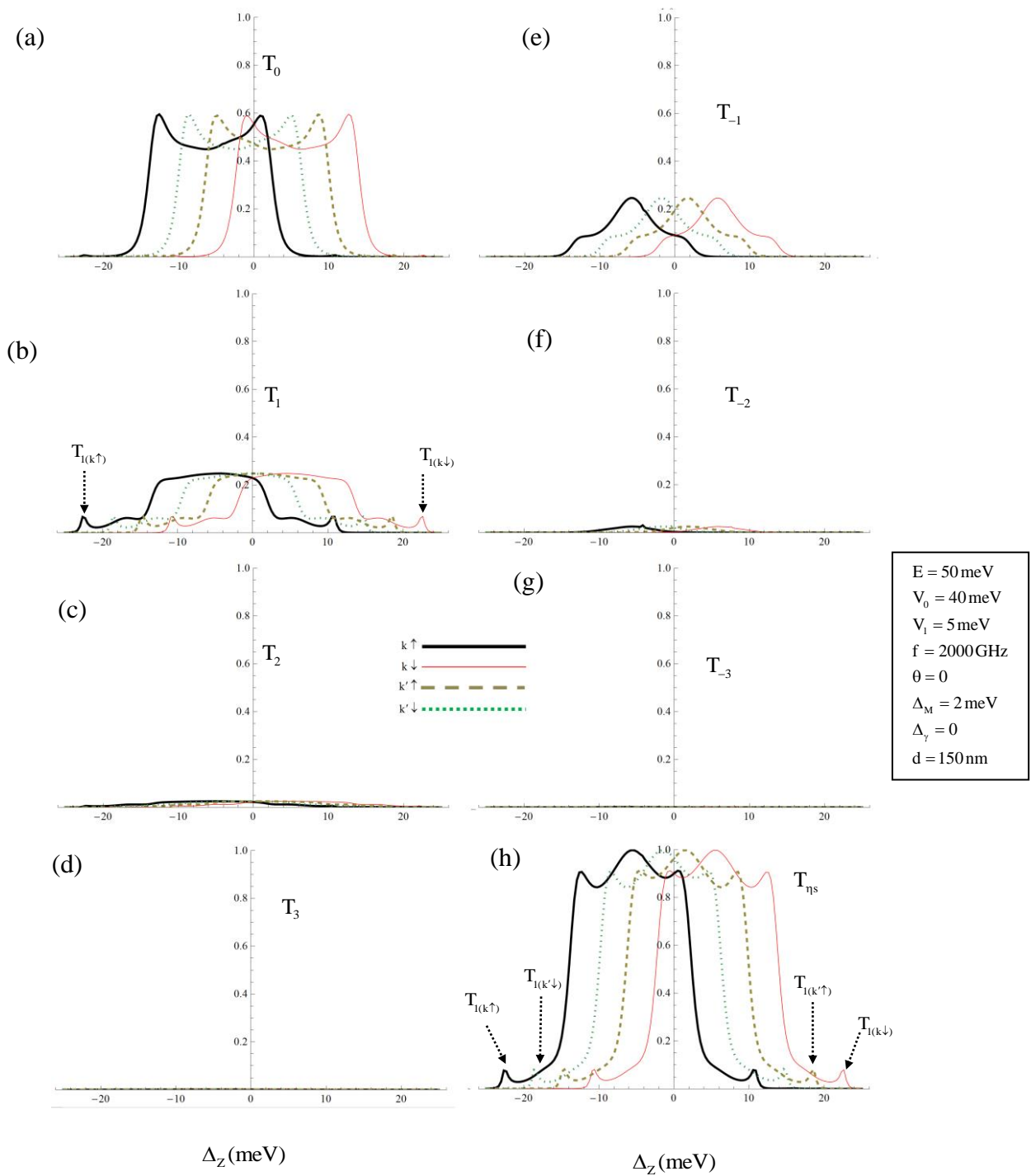

**Figure 4**

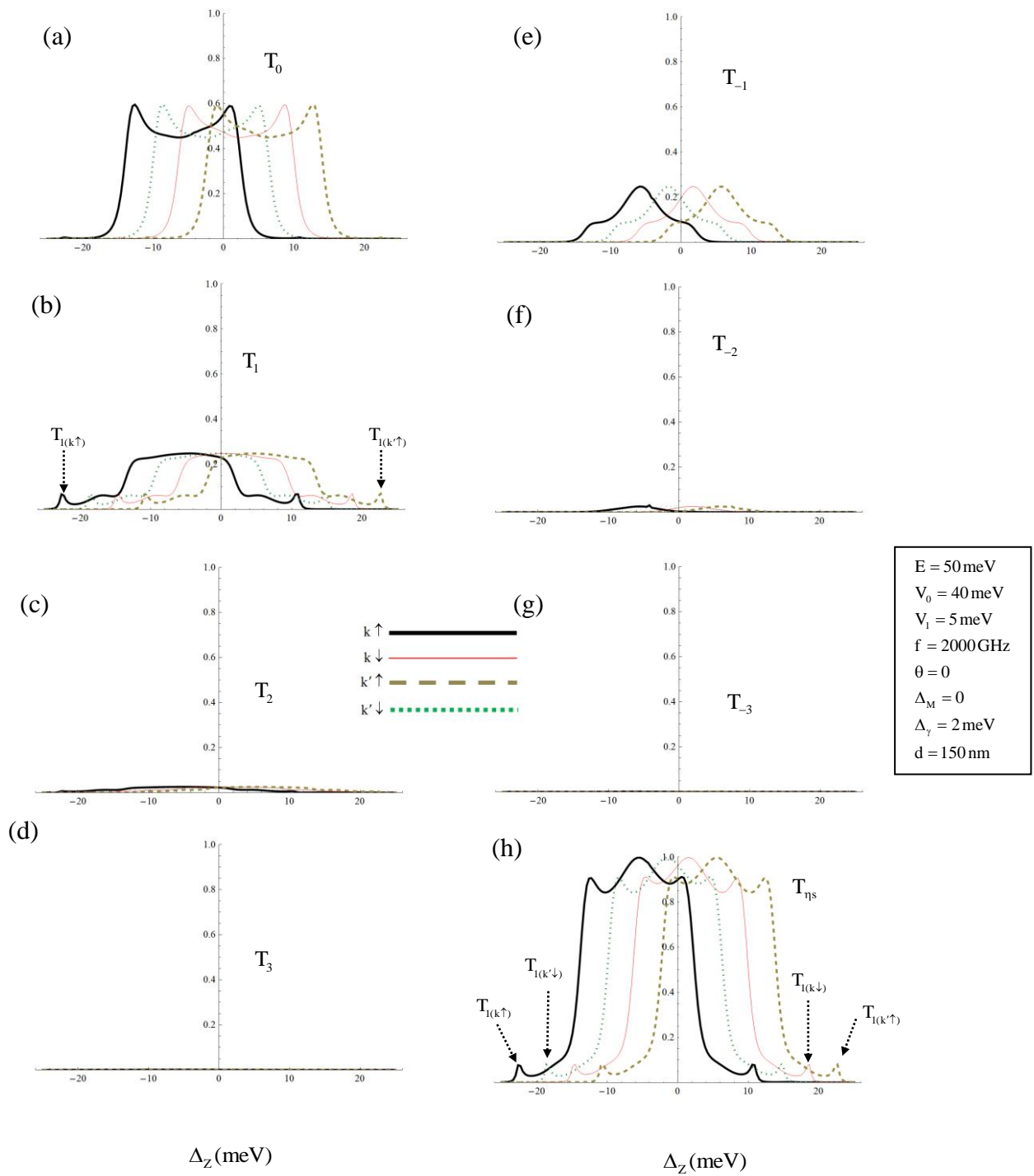

**Figure 5**

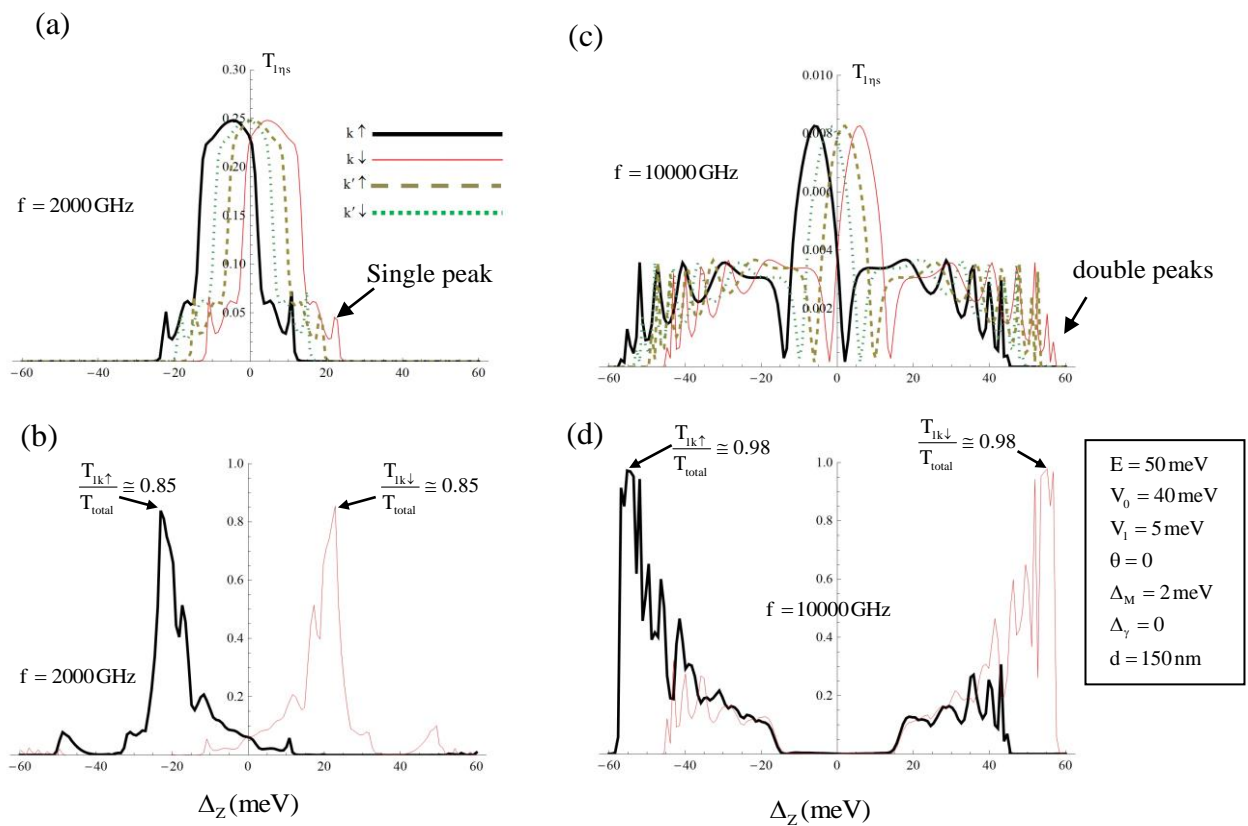

**Figure 6**